\documentstyle[aps]{revtex}

\begin{document}
\draft
\title{ Cyclotron effective mass of 2D electron layer at $\rm \bf
GaAs/Al_xGa_{1-x}As$ heterojunction subject to in-plane magnetic
fields}
\author{L.\ Smr\v{c}ka, P.\ Va\v{s}ek, J.\ Kol\'a\v{c}ek, T.\
Jungwirth, and M.\ Cukr}
\address{Institute of Physics, Academy of Science of
the Czech Republic,\\ Cukrovarnick\'a 10, 162 00 Prague 6, Czech
Republic}
\date{Received \today}
\maketitle

\begin{abstract}
We have found that Fermi contours of a two-dimensional electron
gas at $\rm GaAs/Al_xGa_{1-x}As$ interface deviate from
a standard circular shape under the combined influence of an
approximately triangular confining potential and the strong
in-plane magnetic field. The distortion of a Fermi contour
manifests itself through an increase of the electron effective
cyclotron mass which has been measured by the cyclotron resonance
in the far-infrared transmission spectra and by the thermal
damping of Shubnikov-de Haas oscillations in tilted magnetic
fields with an in-plane component up to 5 T. The observed
increase of the cyclotron effective mass reaches almost 5 \% of
its zero field value which is in good agreement with results of
a self-consistent calculation.
\end{abstract}
%\narrowtext

The electrons confined to the $\rm GaAs/Al_xGa_{1-x}As$
heterostructure form a quasi two-dimensional electron gas. In
a zero magnetic field, or if a magnetic field is applied
perpendicularly to the confinement plane $x-y$, the electron
motion can be separated into an electric contribution governed by
the confining potential $V_{conf}$ and a free in-plane motion
which can be quantized into Landau levels. In this case, the
electron gas can be considered as two-dimensional and the
component of the motion in the third dimension, perpendicularly
to the sample plane, can be neglected.

Subject to an in-plane magnetic field $B_y$, the electrons moving
in the $x$ direction are decelerated or accelerated by the
combined effect of the crossed fields $B_y$ and $ E_z =
-dV_{conf}(z)/dz$, depending on the form of the confining
potential $V_{conf}$, while the electron motion in the $y$
direction remains unmodified. The system no longer behaves as
a strictly two-dimensional one and its third dimension must be
taken into account. As pointed out already in \cite{Cras}
\cite{Zaw}, the subband separation, the 2D density of states and
the shape of the Fermi contour become functions of the magnetic
field $B_y$.

To probe the changes of the electron structure due to the
in-plane magnetic field it is useful to add a perpendicular
component $B_z$ to $B_y$. When the perpendicular component is
weak, the electron dynamic in the $x-y$ plane can be described
semiclassically. In the quasiclassical picture, the electrons
move along real space trajectories with shapes similar to the
shape of the Fermi contour, but multiplied by the scale factor
$\hbar/|e|B_z$ and rotated by $\pi /2$. The \lq egg-like\rq{}
distortion of the Fermi contour and, consequently, of the real
space trajectory results in changes of the cyclotron motion of an
electron which is characterized by the cyclotron frequency
$\omega_c$ and the cyclotron effective mass $m_c$. It is well
known that the quasiclassical cyclotron motion can be quantized
using the standard Bohr-Sommerfeld rule which states that each
real space trajectory encloses an integer number of flux quanta
$\hbar/|e|$. Such procedure yields a discrete spectrum of Landau
levels and the density of states becomes a series of delta
functions separated by $\hbar\omega_c$. The degeneracy of each
level is $2|e|B_z/h$, exactly as in the case of perpendicular
magnetic fields.

The cyclotron effective mass $m_c$ should not be confused with
the electron effective mass which is, for the case of \lq an
egg-like\rq{} Fermi line corresponding to the approximately
triangular well, a tensor and an anisotropic function of the
position on the line. The cyclotron mass is related to the shape
of the Fermi contour by

\begin{equation}
\label{e1}
m_c = \frac{\hbar^2}{2\pi}\;\oint \frac{dk}{|\nabla_kE|}
\end{equation}

\noindent where $dk$ denotes an element of a length of the Fermi
line. Moreover, in the case of two-dimensional systems a simple
relation between the density of states $g$ and the cyclotron mass
of a single subband is obtained,

\begin{equation}
\label{e2}
g = \frac{m_c}{\pi \hbar^2}.
\end{equation}

\noindent Equations (\ref{e1}) and (\ref{e2}) are known for
a long time \cite{Ons,Lif}. Since then they were utilized many
times, for an application to 2D systems see e.g. \cite{Math}.
Their detailed derivation and discussion for the case of combined
parallel and perpendicular magnetic fields can be found in
\cite{Smr}.

This work is devoted to both experimental and theoretical
investigation of the cyclotron effective mass as a function of
the in-plane component of the applied magnetic field. To
determine $m_c$ experimentally, we have measured the cyclotron
resonance in the far-infrared transmission spectra and the
thermal damping of the Shubnikov-de Haas (SdH) oscillations. The
self-consistent calculation of the electronic structure and the
cyclotron effective mass, based on the parameters of a MBE grown
samples, was utilized to obtain the theoretical curves of
$m_c(B_{\parallel})$ dependence.

The samples employed are $\rm GaAs/Al_{x}Ga_{1-x}As$ modulation
doped heterojunctions prepared by molecular beam epitaxy (MBE).
On the top of (100) semi-insulating substrate we have grown first
AlAs/GaAs superlattice, followed by a $\rm 2 \mu m$ undoped GaAs
buffer layer, an undoped 10 nm $\rm GaAs/Al_{0.32}Ga_{0.68}As$
spacer layer, a 100 nm $n$-doped $\rm GaAs/Al_{0.32}Ga_{0.68}As$
layer ($\rm 1.7\times 10^{18} cm^{-3} \, Si$), and finally a 20
nm GaAs cap layer. A~part of a wafer was used to prepare
a rectangular sample $\rm 8\times 8\, mm$ for the cyclotron
resonance experiments, the sample for the magnetoresistance
measurements was shaped into Hall-bar geometry using a standard
lithographic technique. Its length and width were $\rm 1100 \mu
m$ and $\rm 100 \mu m$, respectively. The concentration of 2D
electron gas at a heterojunction, $N_e = \rm 5.2\times 10^{11}
cm^{-2}$, and the mobility $\mu = \rm 4.1\times 10^5 cm^2/Vs$,
were determined from the Shubnikov-de Haas oscillations of the
magnetoresistance and the resistivity, respectively, measured at
a temperature of 4.2 K.

Experimental study of $m_c$ dependence on the in-plane magnetic
field started with the cyclotron resonance measurement. The
employed laser based far-infrared spectrometer is described in
detail elsewhere \cite{Kol}. The sample was placed in the 10
T superconducting solenoid, with the laser beam oriented in
parallel with the coil axis and the magnetic field direction.
A silicon bolometer served to record the radiation transmitted
through the sample as a function of the sweeping magnetic field.
Independent monitoring of the laser power by a pyroelectric
detector, located outside the cryostat, enabled us to eliminate
efficiently random fluctuations of the laser system. A calibrated
Hall probe, located near the sample, was used to determine the
magnetic field.

The cyclotron mass was calculated from the position of
a transmission minimum on the magnetic field scale, according to
an equation

\begin{equation}
\label{e3}
m_c = \frac{|e|B_{\perp} \lambda}{2\pi c},
\end{equation}

\noindent where $B_{\perp}$ denotes the perpendicular component
of a magnetic field for which the resonance occurs, $\lambda$ is
the laser line wavelength and $c$ the light velocity.

In the standard configuration, the sample is oriented with the
plane perpendicular to the magnetic field direction, i.e. there
is no parallel field component. In that case, a laser line with
$\lambda = \rm 129.5 \mu m$ yields the effective mass $m_c =\rm
0.0645$. To determine its field dependence, the sample was
mounted at a tilt, to achieve approximately the same parallel and
perpendicular components of the magnetic field. The angle between
a normal to the sample plane and the field direction was found
$\rm 43^{\circ}47'$ by an optical measurement. For a tilted
sample three different laser lines were employed, with
wavelengths 432.7 $\mu$m, 163.0 $\mu$m and 133.1 $\mu$m, their
transmission curves are shown in figure \ref{f1}. The
corresponding resonance fields have non-zero parallel components
and allow to calculate the increase of the cyclotron mass with
increasing in-plane magnetic field.

The SdH oscillations damping was investigated in the
superconducting magnet reaching the maximum field 7 T. The sample
holder make it possible to fix an arbitrary tilt of a sample with
respect to the applied magnetic field direction. In our case, the
perpendicular configuration and a configuration with the sample
plane almost parallel with the magnetic field were employed. The
Hall voltage, detected on the sample itself, was used as
a measure of the perpendicular field component. The sample was
immersed in the pumped $\rm ^4He$ bath and its temperature
determined from the pressure of the gas.

The interpretation of experimental data is not so straightforward
as in the case of cyclotron resonance. For the perpendicular
magnetic field configuration, the detailed analysis of 2D
electron gas subjected to low/intermediate fields \cite{IS}
showed that the oscillating part of the magnetoresistance is
proportional to the oscillating part of the 2D density of states
broadened by the zero field relaxation time $\tau$. We assume
that this type of behaviour is preserved also in the case of
tilted magnetic fields, only the quantities corresponding to
$B_{\parallel} = 0$ should be replaced by that in $B_{\parallel}
\neq 0$. Then we obtain an expression

\begin{equation}
\label{e4}
\frac{\Delta \varrho}{\varrho_{\circ} } \propto
\frac{\Delta g}{g} = 2 \sum_{s=1}^{\infty} \frac{X}{\sinh X}
\exp(-\frac{\pi s}{\omega_c\tau})
\cos(\frac{2\pi^2 N_e s}{|e|B_{\perp}}-\pi s),
\end{equation}

\noindent where the temperature damping factor $X$ is given by

\begin{equation}
\label{e5}
X = \frac{2\pi^2 k_B T}{\hbar |e|} \frac{m_c}{B_{\perp}}.
\end{equation}

Similarly as in the case of the cyclotron resonance, our study
started with the sample subjected to perpendicular field. In that
case the full Dingle plot can be used to find the effective mass
$m_c$ and the Dingle temperature $T_D = \hbar/2\pi k_B \tau$ with
a high accuracy. The amplitudes of oscillations were determined
from the magnetoresistance curves, after removing their smooth
parts and higher harmonics by a digital filter \cite{Kai}. The
value $m_c = 0.0658$ is in good agreement with $m_c = 0.0645$
obtained from the cyclotron resonance measurement and the widely
accepted value $m_c = 0.067$. The Dingle temperature was equal to
1.31 K.

Since our equipment did not permit to vary $B_{\perp}$ and
$B_{\parallel}$ independently, the magnetoresistance $\Delta
\varrho/\varrho$ was investigated in a tilted magnetic field with
a fixed angle $\rm 80^{\circ} 31'$ between the sample plane and
the field direction. Figure \ref{f2} shows $\Delta
\varrho/\varrho$ obtained with $j \parallel B_{\parallel}$ for
several temperatures and plotted as a function of $B$. In this
case each peak corresponds to a different $B_{\parallel}$ and,
therefore, $m_c(B_{\parallel})$ was determined from the
temperature damping of individual peaks with a limited accuracy,
similarly as in \cite{Har}, where the SdH oscillations in
a double-well structure were investigated. Again, the digital
filter was applied to the data to gain curves comparable with the
first, leading term of equation (\ref{e4}). It was also assumed
that the relaxation time $\tau$ (the Dingle temperature $T_D$)
does not depend on $T$ and that, therefore, the exponential
damping factor in (\ref{e4}) does not influence the temperature
dependence of an amplitude at all.

The figure \ref{f3} presents the relative changes of the
cyclotron effective mass obtained from both the cyclotron
resonance measurement and the damping of SdH oscillation. The
ratio $m_c(B_{\parallel})/ m_c(0)$ is shown, instead of the raw
data, to suppress possible systematic errors introduced into
the results by uncertainties of magnetic field calibration and
angle measurement in two experimental arrangements. The overall
increase in $m_c$ is approximately 5\% in the maximum parallel
field 5 T and there is quite acceptable agreement between the
changes determined from the cyclotron resonance and SdH
measurements. Together with the experimental data also the
theoretical curves are shown.

Before discussing to what extend the theory agrees with the
experimental data, we briefly describe the employed computational
method. We have used a standard semi-empirical model working
quantitatively for the lowest conduction states of $\rm
GaAs/Al_{x}Ga_{1-x}As$ heterostructures. The coupled Poisson and
Schr\"{o}dinger equations are solved, the Schr\"{o}dinger
equation is treated in the envelope function approximation and
envelope functions are assumed to be built from host quantum
states belonging to a single parabolic band. Since the effect of
the effective mass mismatch is completely neglected, the envelope
functions of GaAs and $\rm Al_{x}Ga_{1-x}As$ are smoothly matched
at the interface. The Hartree term of the confining potential is
determined from the Poisson equation and we use the local
density-functional approximation \cite{RD} of the exchange
correlation term.

The input parameters for calculations were taken from our
knowledge of samples described above. The conduction band offset
$V_b$ was determined from the AlAs fraction $x$ according to
a semi-empirical rule

\begin{equation}
\label{e6}
 V_b = \alpha \times 1.247 x \,\,\, \rm(eV),
\end{equation}

\noindent with the partition ratio $\alpha$ in the range
0.60-0.65 \cite{Hi}, the ionization energy of Si donors was
taken $\rm 60 meV$. The concentration of donors was $N_d = \rm
1.7\times 10^{18} cm^{-3}$. As for the acceptors, their energy
levels are assumed to lie 1.514 eV below the conduction band
bottom and the dielectric constant $\varepsilon =\rm 12.9$ was
employed.

Similarly as in \cite{St}, we have found in the course of
calculations that the results are relatively insensitive to the
exact values of most of these parameters, except of the
concentration of residual acceptors in GaAs. Unfortunately, this
quantity is difficult to measure and there is almost no
information available concerning the homogeneity of their
distribution in the $z$ direction. Here we accepted the usual
approximation of constant ionized impurities concentration
$N^{-}_a(z) = N_a$ for $0 \leq z \leq l_a$, with the parameter
$l_a$ determined from the conditions of charge neutrality and the
thermodynamical equilibrium in each loop of the self-consistent
procedure. A series of different $N_a$ has been tested. For
presentation in figure \ref{f3}, we have chosen two different
\lq realistic\rq{} values corresponding to the standard level of
impurities in our MBE samples , $N_a = \rm 1.3\times 10^{14}
cm^{-3}$ and $N_a = \rm 2.3\times 10^{14} cm^{-3}$, and slightly
modified $\alpha$ in equation (\ref{e6}), in the given range, to
preserve the concentration $N_e$ of free electrons equal to the
experimentally determined value $\rm 5.2\times 10^{11} cm^{-2}$.

Having in mind the above note about a semi-empirical nature of
the self-consistent calculations, we can conclude that there is
a good agreement between the theoretically predicted field
dependence of the cyclotron effective mass and the experimental
data. We believe that the above described experimental methods
might be appropriate for investigating the rich variety of Fermi
contours in quantum wells with diverse shapes of confining
potential and in wider range of in-plane magnetic fields.

This work has been supported by the Academy of Science of the
Czech Republic under Contracts No. 110 423 and No. 110 414, and
by NSF, U.\ S., through the Grant NSF INT-9106888.

\begin{figure}
\caption{Selected curves of the transmission coefficient of the
infrared radiation through a tilted sample. The angle between the
beam and the normal to the sample plane was $\alpha = \rm
43^{\circ}47'$ and $T \approx \rm 2 K$.}
\label{f1}
\end{figure}

\begin{figure}
\caption{ Examples of magnetoresistance curves for a tilted
sample, with the angle $\alpha =\rm 9^{\circ}30'$, for the
temperatures 2.70 K, 2.88 K, 3.15 K, 3.41 K, 3.59 K and 3.82 K.
Smooth parts of curves slightly increase with growing T, while
the oscillations amplitudes are damped.}
\label{f2}
\end{figure}

\begin{figure}
\caption{ Relative changes of the cyclotron effective mass $m_c$
obtained from the cyclotron resonance measurement (circles) and
the damping of SdH oscillation (triangles). The upper line is the
theoretical curve calculated for the concentration of acceptors
$N_a = \rm 1.3\times 10^{14} cm^{-3}$, the second line for $N_a
= \rm 2.3\times 10^{14} cm^{-3}$. The inset demonstrates the
variation of the Fermi line, corresponding to the lower
concentration, caused by the applied in-plane magnetic field
$B_{\parallel}$.}
\label{f3}
\end{figure}


\begin{thebibliography}{99}
\bibitem{Cras} J.\ H.\ Crasemann, U.\ Merkt and J.\ P.\ Kotthaus,
Phys.\ Rev.\ B {\bf 28}, 2271 (1983).

\bibitem{Zaw} W.\ Zawadzki, S.\ Klahn and U.\ Merkt,
Phys.\ Rev.\ B {\bf 33}, 6916 (1986).

\bibitem{Ons} L.\ Onsager, Phil.\ Mag. {\bf 43}, 1006 (1952).

\bibitem{Lif} I.\ M.\ Lifshitz, Sov.\ Phys. - JETP {\bf 30}, 63
(1956).

\bibitem{Math} T.\ G.\ Matheson and R.\ J.\ Higgins, Phys.\ Rev.\
B {\bf25}, 2633 (1982).

\bibitem{Smr} L.\ Smr\v{c}ka and T.\ Jungwirth, J.\ Phys.: Condens.
Matter {\bf 6}, 55 (1994).

\bibitem{Kol} J.\ Kol\'a\v{c}ek, Z.\ \v{S}im\v{s}a and R.\ Tesa\v{r}, Meas.\
Sci.\
Technol.\ {\bf 4}, 1085 (1993).

\bibitem{IS} A.\ Isihara and L.\ Smr\v{c}ka, J.\ Phys.\ C {\bf 19},
6777 (1986).

\bibitem{Kai} J.\ F.\ Kaiser and W.\ A.\ Reed, Rev.\ Sci.\
Instrum.\ {\bf 19}, 1103 (1978).

\bibitem{Har} N.\ E.\ Harff, J.\ A.\ Simmons, S.\ K.\ Lyo, J.\
F.\ Klem and S.\ M.\ Goodnick, {\it ICPS-22 Proceedings}, in
press, S.\ K.\ Lyo, preprint.

\bibitem{RD} P.\ Ruden and G.\ H.\ D\"{o}hler, Phys.\ Rev.\ B {\bf
27}, 3538 (1983).

\bibitem{Hi} S.\ Hiyamizu, in {\it Semiconductors and Semimetals,
\bf 30}, edited by R.\ K.\ Willardson and A.\ C.\ Beer (Academic
Press, Boston, 1990) p. 75.

\bibitem{St} F.\ Stern and S.\ Das Sarma, Phys.\ Rev.\ B {\bf
30}, 840 (1984).

\end{thebibliography}
\end{document}